\documentclass[a4paper,fleqn,usenatbib]{mnras}

\usepackage[T1]{fontenc}
\usepackage{ae,aecompl}

\usepackage{subfig}

\usepackage{graphicx}	
\usepackage{amsmath}	
\usepackage{amssymb}	
\usepackage{sidecap}

\usepackage{caption}

\title[Formation of C-molecules on Graphitic Surfaces]{Simulating the Formation of Carbon-rich Molecules on an idealised Graphitic Surface}

\author[D. W. Marshall and H. R. Sadeghpour]{
David W. Marshall,$^{1,2,3}$\thanks{E-mail: marshall@mps.mpg.de}
and H. R. Sadeghpour$^{3}$
\\
$^{1}$Max-Planck-Institut f{\"u}r Sonnensystemforschung, Justus-von-Liebig-Weg 3, 37077, G{\"o}ttingen, Germany \\
$^{2}$School of Physics and Astronomy, University of Southampton, Highfield, Southampton, SO17 1BJ, UK \\
$^{3}$ITAMP, Harvard-Smithsonian Center for Astrophysics, Cambridge, MA 02138, USA
}

\date{Accepted XXX. Received YYY; in original form ZZZ}

\pubyear{2015}

\begin{document}
\label{firstpage}
\pagerange{\pageref{firstpage}--\pageref{lastpage}}
\maketitle

\begin{abstract}
There is accumulating evidence for the presence of complex molecules, including carbon-bearing and organic molecules, in the interstellar medium. Much of this evidence comes to us from studies of chemical composition, photo- and mass-spectroscopy in cometary, meteoritic and asteroid samples, indicating a need to better understand  the surface chemistry of astrophysical objects. There is also considerable interest in the origins of life-forming and life-sustaining molecules on Earth. Here, we perform reactive molecular dynamics simulations to probe the formation of carbon-rich molecules and clusters on carbonaceous surfaces resembling dust grains and meteoroids. Our results show that large chains form on graphitic surfaces at low temperatures (100K - 500K) and smaller fullerene-like molecules form at higher temperatures (2000K - 3000K). The formation is faster on the surface than in the gas at low temperatures but slower at high temperatures as surface interactions prevent small clusters from coagulation. We find that for efficient formation of molecular complexity, mobility about the surface is important and helps to build larger carbon chains on the surface than in the gas phase at low temperatures. Finally, we show that the temperature of the surface strongly determines what kind of structures forms and that low turbulent environments are needed for efficient formation.
\end{abstract}

\begin{keywords}
(ISM:) dust, extinction < Interstellar Medium (ISM), Nebulae -- ISM: molecules < Interstellar Medium (ISM), Nebulae -- meteorites, meteors, meteoroids < Planetary Systems
\end{keywords}

\section{Introduction}

A diverse variety of complex molecules have been observed from a range of astronomical bodies, such as planetary nebulae \citep{cami2010detection}, dark clouds \citep{friberg1988methanol}, hot cloud cores \citep{cazaux2003hot,van2003sulphur}, meteorites \citep{botta2002extraterrestrial, plows2003evidence}, AGB outflows and circumstellar envelopes \citep{tielens2008interstellar}. The list of molecules is long but it includes ethene, benzene, methanol and propynal \citep{herbst2009complex}, as well as acids \citep{cazaux2003hot}, amino acids \citep{botta2002extraterrestrial} and alcohols \citep{friberg1988methanol}. Spectroscopic observations have also been made of oxide and carbonaceous dusts like amorphous silicates, crystalline forsterite, graphite and silicon carbide \citep{tielens2011chemical,draine2003interstellar}. 

Recently, complex cyanides have been detected in a protoplanetary disk around the young star MWC 480. The presence of these cyanides can be attributed to molecular processing in gas phase and grain surface interactions which create molecules of greater complexity. Crucially, they have been found in similar abundances to that of comets in our solar system \citep{oberg2015comet}. The similarities between the abundances of the nitrogen bearing organic molecules in these objects suggests that the profusion of carbon chemistry in our solar system is not unique and may also be found elsewhere. 

Broad emission features have been revealed in the mid-infrared spectrum in observations by the Infrared Space Observatory (ISO), the Spitzer Space Observatory (SSO), and ground and airborne based studies. The progenitors for these lines have been attributed, in ubiquity in astrophysical objects ranging from the surface of dark clouds, reflection and planetary nebulae, HII regions, nuclei of galaxies, to the infrared fluorescence of far-ultraviolet pumped polycyclic aromatic hydrocarbons (PAHs) with around 50 atoms \citep{tielens2008interstellar}. 
Another particularly important species of organic molecule are fullerenes, also known as buckyballs \citep{tielens2008interstellar}. Fullerenes are spherical molecules of pure carbon. They have rings of five and six atoms, and odd numbered rings are responsible for structural curvature. Examples include C$_{20}$, C$_{60}$, C$_{70}$ and C$_{240}$. 

It is thought that about 10\% of the carbon in the universe is locked in PAHs and small carbonaceous grains such as fullerenes \citep{herbst2009complex}. The carbon locked inside these molecules could be the biological precursors to the formation of amino acids and proteins, and hence important to the origins of life. Laboratory experiments \citep{chen2008amino} have shown that UV irradiation of naphthalene in an ice mixture can lead to the production of an organic residue containing many identifiable amino acids. 

Fullerenes are the largest molecules observed in space and their formation mechanism will shed light on how complex molecules may be produced \citep{bernard2014interstellar}. In contrast to PAHs, fullerenes are stable against irradiation and can hence survive harsh interstellar environments. The surface chemistry of fullerenes, perhaps through the formation of aliphatic chains, could be interesting for the production of organic molecules. Fullerenes certainly have the ability to adsorb atoms and molecules onto their surface and interact with them due to its properties as an electron acceptor.

A host of discoveries on the origins of organic and carbonaceous matter and delivery to Earth, D/H ratio in water \citep{hartogh2011ocean,morbidelli2000source,altwegg201567p,hassig2015time}, and presence of PAH precursor molecules in chondrites \citep{plows2003evidence,engrand1998carbonaceous}, and the low temperatures of the C$_{60}$ vibrational lines \citep{cami2010detection}, point to synthesis and deposition in bulk and on astrophysical surfaces. It is therefore important to investigate the surface chemistry of astrophysical objects for the production of organic molecules. The aim of this project is to use reactive molecular dynamics simulations to probe the formation of carbon-rich molecules and clusters on carbonaceous surfaces resembling dust grains and meteoroids. In simulations performed by \citep{patra2014nucleation}, at temperatures ranging from 500-3000K, it was found that at high temperatures (2000-3000K), fullerene-like structures formed, and clusters and curved features appeared at early times. In contrast, at lower temperatures (500-1000K), long chain structures preferentially formed and molecular curvature only occurred at later times. Patra et al produced population distributions of the results, binning clusters according to their size. These distributions showed that larger molecules form at lower temperatures as high thermal fluctuations inhibit the growth of big molecules at higher temperatures. We will study contrast and departure from what's been inferred in the gaseous phase and learn if surfaces can assist in catalysis of large carbon chain or cage formation.

This paper contains three further sections: in Section 2, we describe the computational molecular dynamics method used in our simulations as well as some limitations of our system; in Section 3, our results are presented and the astronomical significance of the findings are discussed; and in Section 4, we summarise the main results in a conclusion.  

\section{Computational Method}

We use the Large Atomic/Molecular Massively Parallel Simulator (LAMMPS) \citep{plimpton1995fast} to perform the reactive molecular dynamics calculations and compute the potentials, interactions and trajectories of the particles in our system. The results are then visualised using Visual Molecular Dynamics software (VMD) \citep{humphrey1996vmd}. Details are provided below.

\subsection{Molecular dynamics}

Molecular dynamics (MD) simulations give a classical description of the movement of a large number of particles in a system. They incorporate two-body and three body-interactions through density functional theory (DFT) calculations using quantum mechanical or empirical potentials. So that we can obtain coarse-grain information on formation mechanisms on surfaces from large-atom number simulations, we resort to the use of empirical potentials. 

The MD calculations start by initialising the system with the particle position and velocity. The particle velocity is scaled to the mean kinetic energy determined from the initial temperature. The force which acts on each particle is then computed from a potential which describes pairwise interactions and incorporates three-body recombination to form dimers. From this, Newtons equations of motion can be solved and the trajectories of the particles are found.

 An empirical AIREBO (Adaptive Intermolecular Reactive Empirical Bond Order) potential \citep{stuart2000reactive} is used to model the two-body and reactive interactions between the atoms. The potential function takes the form:
\\
\textbf{
\begin{equation}
E^{AIREBO}=  E^{LJ} + E^{tors} + E^{REBO}
\label{eq:airebo}
\end{equation}
\begin{equation}
E^{LJ}_{ij} =  4 \epsilon _{ij} S \Bigg[ \Bigg( \frac{\sigma_{ij}}{r_{ij}} \Bigg)^{12} - \Bigg( \frac{\sigma_{ij}}{r_{ij}} \Bigg)^{6} \Bigg]
\label{eq:lj}
\end{equation}
\begin{equation}
E^{REBO}_{ij} = V^{R}(r_{ij}) + b_{ij}V^{A}(r_{ij})
\label{eq:rebo}
\end{equation}
}
\\

The AIREBO potential in Equation~\ref{eq:airebo} has three terms: a Leonard Jones 12-6 term, E$^{LJ}$; a torsional term, E$^{tors}$; and a REBO (Reactive Empirical Bond Order) term, E$^{REBO}$. The Leonard-Jones 12-6 part in Equation~\ref{eq:lj} accounts for the long range attraction between particles $( \frac{\sigma_{ij}}{r_{ij}})^{6}$ and the short range repulsion $( \frac{\sigma_{ij}}{r_{ij}})^{12}$. The range extends out to a cut-off distance determined by the type of interacting pairs (for example, for a C-C pair this is 10.2 \AA ) and the switching function $S$ within E$^{LJ}$ turns off this potential at short distances so that it does not interfere with the behaviour of the REBO potential; $\epsilon_{ij}$ and $\sigma_{ij}$ are parameters of the atoms involved in the interaction. The torsional part is an explicit four-body potential that accounts for the rotation of molecules about single bonds. This is vitally important when looking at large molecular structures. It arises from the local coordination of atoms and depends upon the dihedral angle made in various hydrocarbon configurations.

The REBO term in Equations~\ref{eq:airebo} and \ref{eq:rebo}, is based upon the Brenner bond-order potential \citep{brenner1990empirical} which permits hybridisation: the formation and breaking of chemical covalent bonds. V$^{R}_{ij}$ and V$^{A}_{ij}$ are respectively, the repulsive and attractive pair potentials between atoms i and j, separated by r$_{ij}$. 
Equation~\ref{eq:rebo} also has a bond order term, b$_{ij}$, from a Tersoff potential \citep{tersoff1988new}. This many-body term varies the strength of the bonding between i and j atoms based upon the co-ordination of the local system. It does this by accounting for the coordination numbers, bond angles and conjugation effects of nearby atoms, thus controlling the many-body interactions.

This potential has been widely used in hydrocarbon systems to model the chemical and intermolecular interactions, for example, in carbon nanotubes, \citep{volkov2010structural}, graphite \citep{ito2012molecular} and organic polymers \citep{yamada2004molecular}.

\subsection{Simulation box with a monolayer surface}

A graphitic surface is formed to sit at the bottom of the simulation box in the z=0 plane. This is done by designing a unit cell with periodic boundary conditions. For the purposes of this work, we will use graphene with a lattice constant of 2.46 \AA ~\citep{fuchs2008introduction}. It should be made clear that this is an idealisation of a surface and is merely intended to emulate a carbonaceous object. Using graphene does have computational advantages, as it is only one atom thick and so reduces the number of atoms needed in the system. There are 4128 carbon atoms in the graphene surface with dimensions 102.5 \AA~x 106 \AA. The simulations are run with temperatures up to 3000K so it is important to note that the graphene surface will still be stable in such conditions.

Once the surface has formed in the simulation box, carbon atoms can be randomly distributed above the graphene and the MD process begins. The system is briefly minimised for 0.02 ps. Constant NVT integration updates the position and velocity of the particles, with Langevin thermostats \citep{schneider1978molecular} thermalising the system to keep the gas and surface atoms at the desired temperature. We use a timestep of 0.5 fs. The system is in the canonical ensemble in non-thermodynamic equilibrium. The box is periodic in the x- and y-directions and has a volume of approximately 10$^{6}$ \AA$^{3}$.

\subsection{Computational limitations}

The simulations are computationally intensive. They are performed on the Odyssey research cluster at Harvard University, which has 60,000 CPUs and 2,140 nodes. It takes about one day of run time to perform the calculations at 500K over a 1 ns time period on a 64 core machine with 4328 atoms. The simulation box is divided into an 8$\times$8$\times$1 grid to ensure an even load balance between the CPUs. Through speed testing, utilising 64 cores on a single node was found to be the optimal configuration. Using more cores, means using additional nodes which creates a lag time problem as the nodes communicate, decreasing the availability of resources and increasing the job queuing time, so often 32 cores are used rather than 64. 

Further computational problems arise due to the fact that the run time is increased as the temperature is raised or as more atoms are included. Even a small graphene surface, like the one we use here (area = 10$^{4}$ \AA $^{2}$), contains 4128 atoms. It is therefore necessary to use a high particle density in order to produce results quickly and avoid small number statistics. In our simulations, this takes a value of $\sim$ 10$^{-4}$ \AA$^{-3}$ {($\sim$ 10$^{20}$ cm$^{-3}$)} which is many orders of magnitude higher than what is seen in typical ISM environments. This is unavoidable, but in spite of these limitations, the results obtained here point to processes occurring in low density astrophysical conditions.

The majority of the results presented here are run over a few nanoseconds which is a sensible period as carbon structures have been shown to form on picosecond timescales \citep{irle2006theory, irle2003formation}. The results in Section~\ref{sec:morph} took several months of run time to complete and represent about $0.4-0.7 \mu s$ of simulation time.

\section{Results and Discussion}

In this section, we present the results of our simulations across a range of temperatures and times. Calculations are run with different numbers of atoms in the system and at different densities. The effect of changing the number of graphene layers is explored for the emergence of bulk behaviour. We also look at how an inflow onto the surface affects the formation of carbon molecules. In all simulations, only carbon atoms are considered in the gas phase.

\subsection{Cluster morphology}
\label{sec:morph}

First, we investigate the morphology of the carbon clusters formed on the graphene surface on long timescales and over a range of temperatures. We randomly place 200 carbon atoms in the simulation box above the surface. In all of the discussions below, the surface and gas atoms are kept at the same temperature. We will discuss the effect of a temperature gradient between the graphene and gas atoms later in Section~\ref{sec:tempgrad}.

The bonding in graphene is covalent and strong. Each carbon atom forms three sp$^{2}$ hybridised orbitals which overlap neighbouring orbitals to create sigma bonds and the distinctive hexagonal shape \citep{schabel1992energetics}. It can also be said to be aromatic although its aromaticity is different from the aromaticity in benzene, coronene, or circumcoronene \citep{popov2012graphene}. Graphene has two $\pi$-electrons in each hexagonal ring which add to its strength. This overlapping of orbitals makes incredibly strong covalent carbon-carbon bonds and gives graphene a high melting temperature; estimated to be about 4900K, based on Monte Carlo simulations \citep{zakharchenko2011melting}. This is higher than the melting temperature found for similar carbon structures such as fullerenes and carbon nanotubes, with values of 4000K and 4800K, respectively. 

\begin{figure*}
\begin{tabular}{cccc}
\subfloat[100K, surface]{\includegraphics[width=0.6\columnwidth]{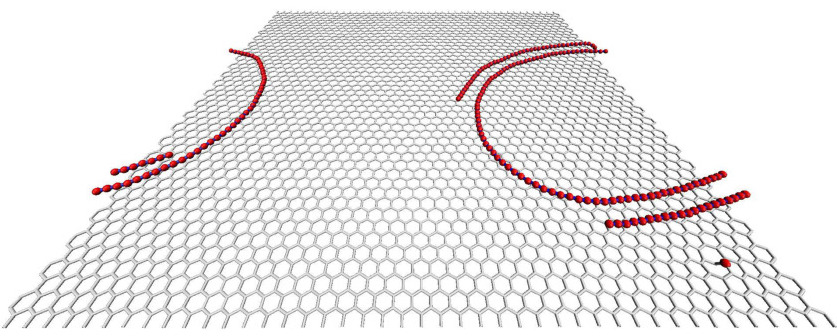}}&
\subfloat[100K, gas]{\includegraphics[width=0.3\columnwidth]{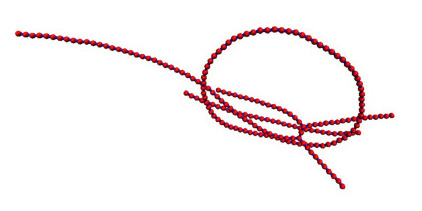}}&
\subfloat[500K, surface]{\includegraphics[width=0.6\columnwidth]{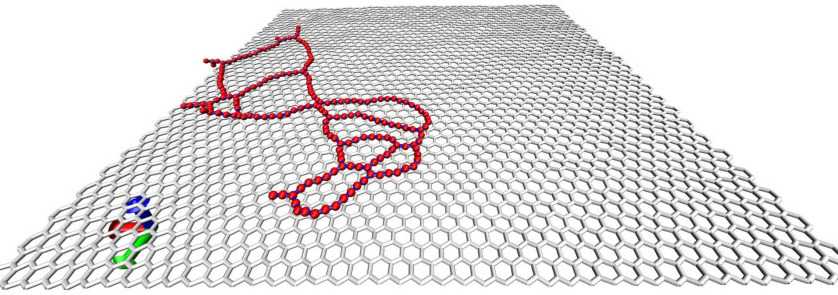}}&
\subfloat[500K, gas]{\includegraphics[width=0.3\columnwidth]{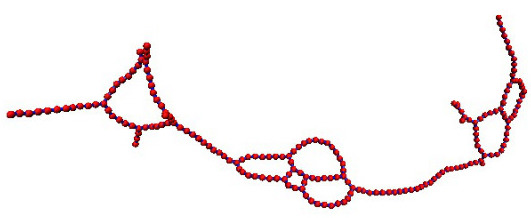}}\\
\subfloat[1000K, surface]{\includegraphics[width=0.6\columnwidth]{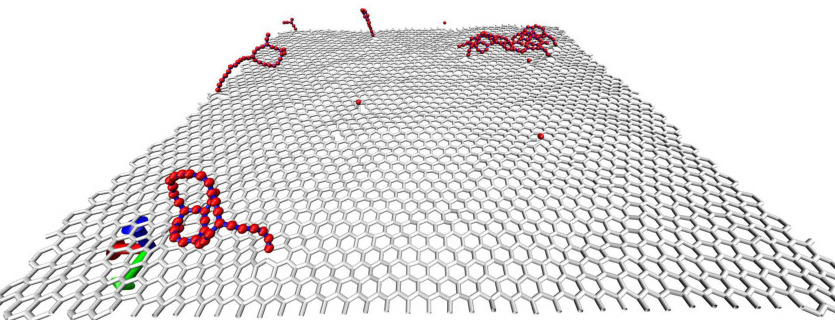}}&
\subfloat[1000K, gas]{\includegraphics[width=0.3\columnwidth]{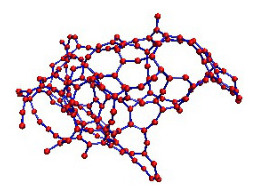}}&
\subfloat[1500K, surface]{\includegraphics[width=0.6\columnwidth]{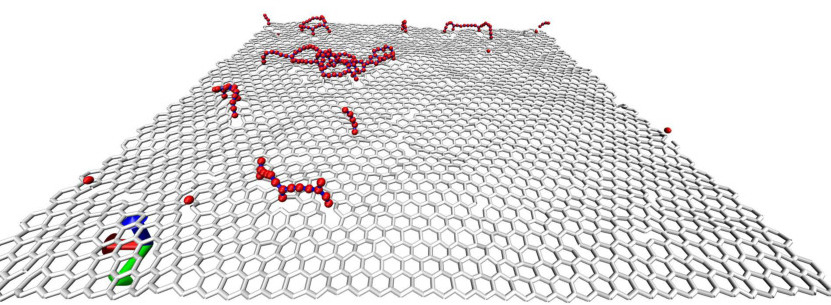}}&
\subfloat[1500K, gas]{\includegraphics[width=0.3\columnwidth]{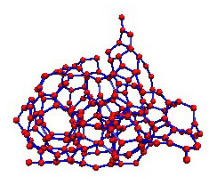}}\\
\subfloat[2000K, surface]{\includegraphics[width=0.6\columnwidth]{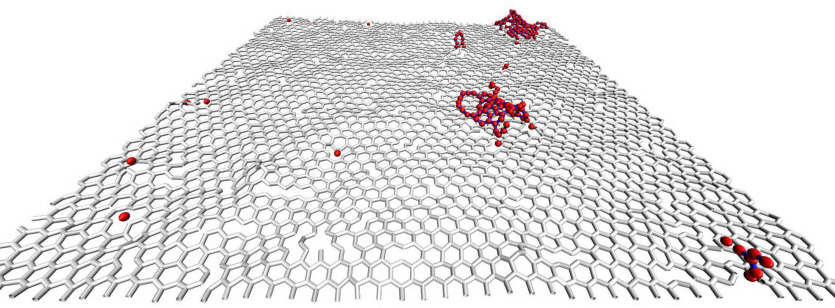}}&
\subfloat[2000K, gas]{\includegraphics[width=0.3\columnwidth]{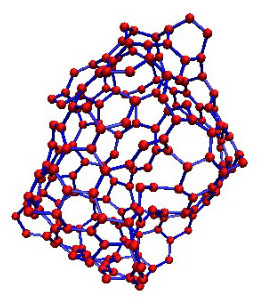}}&
\subfloat[2500K, surface]{\includegraphics[width=0.6\columnwidth]{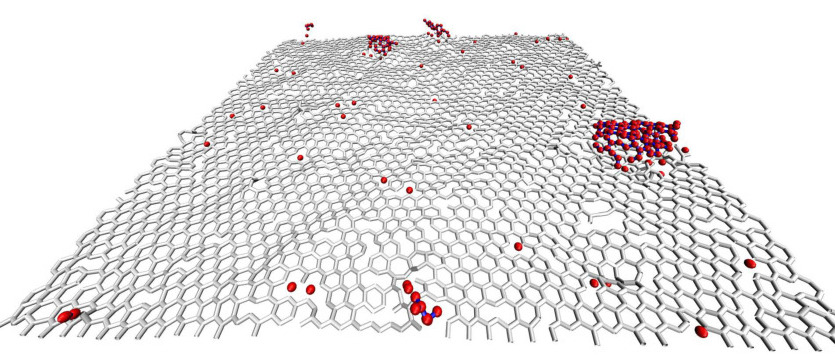}}&
\subfloat[2500K, gas]{\includegraphics[width=0.3\columnwidth]{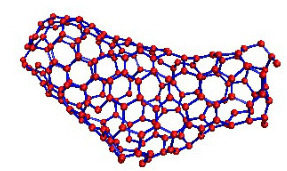}}\\
\subfloat[3000K, surface]{\includegraphics[width=0.6\columnwidth]{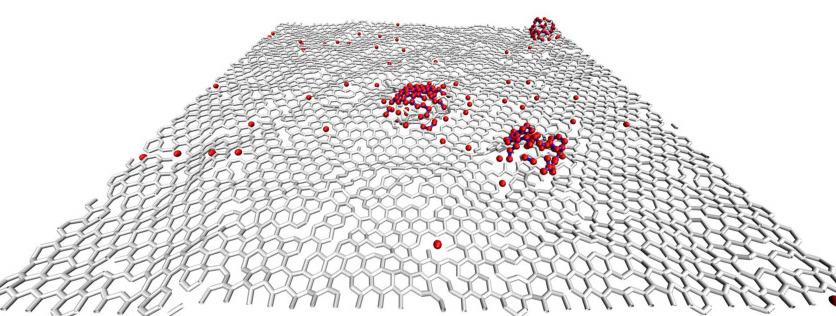}}&
\subfloat[3000K, gas]{\includegraphics[width=0.3\columnwidth]{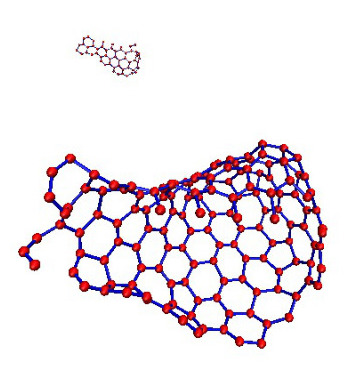}}
\end{tabular}
\caption[Cluster formation on the surface and in the gas phase after 100 ns]{Cluster formation on the surface and in the gas phase after 100 ns from 100K to 3000K. The carbon atoms originating in the gas phase are shown in red, and the bonds between them in blue. The bonds of the graphene surface are depicted in white. At 100K and 500K, there are carbon chains on the surface and in the gas but at higher temperatures (1000K - 2000K) there are irregular clusters. At the 2500K and 3000K, there are fullerene-like carbon cages.}
\label{fig:snap100}

\end{figure*}

Our first set of results (Figure~\ref{fig:snap100}) show the formation of clusters after 100 ns in temperatures ranging from 100 - 3000K. At 100K, two chains form on the surface: one in a spiral and the other following the curvature of the first. The chains can move around but remain at about 3.45 \AA~above the surface. In the gas phase, there is one carbon loop and two chains. In both cases, the chains are bundled closely together but not joined, influenced by the attraction of the Leonard Jones potential. The chains do not have sufficient energy to bind. At 500K, a single flat mulit-ring carbon molecule containing 197 atoms forms on the graphene. Again, it can move about on the surface with a separation of about 3.54 \AA. The carbon chain in the gas phase has similar large ring features but has a filament morphology. This is a result of the periodic nature of the simulation box: the molecule has joined onto itself at both ends, thus stretching it out.

Similar features can be seen at 1000K, 1500K and 2000K. In all three cases, several isolated irregular clusters form which are bound directly onto the surface. They exhibit tighter carbon rings than at colder temperatures, protruding normal to the surface. This effect is due to thermal fluctuations of the carbon molecules at higher temperatures. In the gas phase, we see an irregular carbon cage-like structure forming at each temperature. They are not quite spherical in that there are aliphatic carbon chains coming off them and the constituent carbon rings get tighter as the temperature increases towards the five- and six-member rings of fullerenes.

At 2500K, there are again isolated clusters on the graphene but they are far more regular. In fact one cluster has organised itself into a graphene-like flake with hexagonal rings but no pentagonal rings. This regularity is also exhibited in the gas, where a smoothly curved carbon cage has formed. 

\begin{figure}
\begin{center}
\includegraphics[width=\columnwidth]{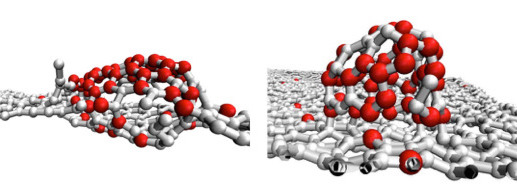}
\caption[VMD snapshots: spherical molecules at 3000K after 100 ns]{Spherical molecular structures forming on the graphene surface at 3000K after 100 ns. They are similar in shape to fullerenes. The red atoms originated in the gas phase and the white atoms are from the surface.}
\label{fig:snapsph100}
\end{center}
\end{figure}

Similar behaviour is seen at 3000K, with regular clusters forming on the graphene surface and in the gas phase, containing hexagonal and pentagonal carbon rings. The pentagonal rings are defects which are essential to forming fullerenes. In fact, some interesting and notable features grow on the surface, as portrayed in Figure~\ref{fig:snapsph100}. The surface clusters bend and deform the surface by overcoming the thermal barrier to help create spherical molecular structures resembling buckyballs.

In observations made by \citep{cami2010detection} in Tc-1, a young planetary nebula around a white dwarf, the PN spectra contained vibrational lines of C$_{60}$ and C$_{70}$ fullerenes. Cami et al argued that these molecules were bound to the surface of a dust grain rather than in the gas phase due to the symmetry of the spectral lines. The photon energies from the planetary nebula imply that large gas-phase species should be at 800 - 1000K but the fullerene spectral lines suggest that they are at a much lower temperature (180 - 330K). For this to be the case, the fullerenes must be in contact with a cool solid surface and the most likely explanation is carbonaceous dust grains in the outflow. We are able to simulate the growth of fullerene-like molecules on surfaces, but the surface temperature is an order of magnitude different than in the observations.

\begin{figure}
\begin{center}
\includegraphics[width=\columnwidth]{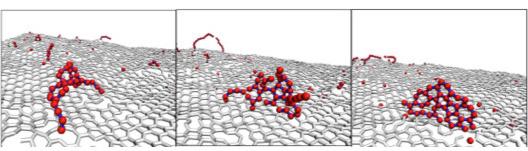}
\caption[VMD snapshots: graphene-like flake at 3000K after 4 ns]{Formation of a graphene-like flake on the surface at 3000K. The snapshots are taken after 0.5, 2 and 4 ns; from left to right. The atoms in the irregular cluster rearrange themselves into a flake with rings of 5, 6 and 7 atoms.}
\label{fig:flake}
\end{center}
\end{figure}

Figure~\ref{fig:flake} shows some further interesting behaviour at 3000K. It depicts an irregular cluster forming in the gas phase and bonding to the surface after 0.5 ns. Over the next 4 ns, the cluster rearranges itself into a regular graphene-like flake with carbon rings of five, six and seven atoms. This shows how easily polycyclic molecules can form at high temperatures.

\begin{figure*}
\begin{tabular}{cccc}
\subfloat[500K, surface, 915 ns]{\includegraphics[width=0.6\columnwidth]{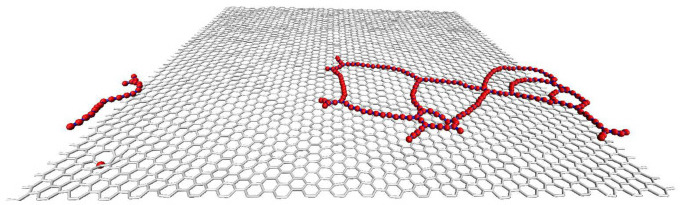}}&
\subfloat[500K, gas]{\includegraphics[width=0.3\columnwidth]{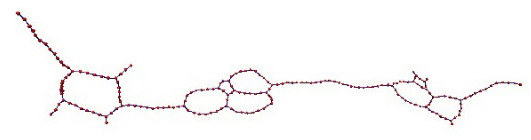}}&
\subfloat[1000K, surface, 883 ns]{\includegraphics[width=0.6\columnwidth]{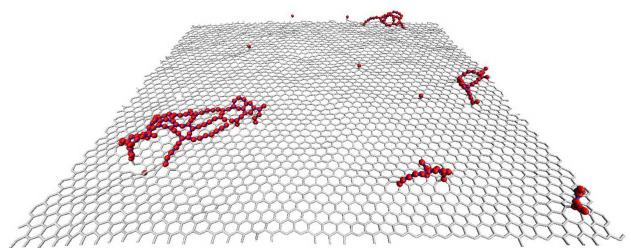}}&
\subfloat[1000K, gas]{\includegraphics[width=0.3\columnwidth]{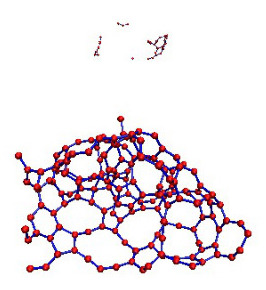}}\\
\subfloat[1500K, surface, 721 ns]{\includegraphics[width=0.6\columnwidth]{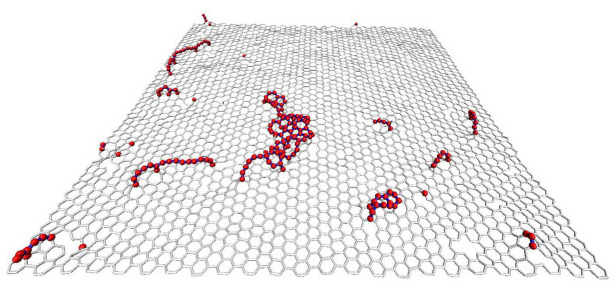}}&
\subfloat[1500K, gas]{\includegraphics[width=0.3\columnwidth]{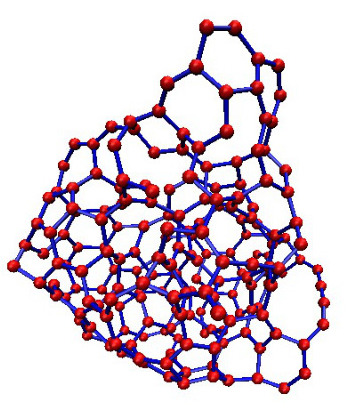}}&
\subfloat[2000K, surface, 644 ns]{\includegraphics[width=0.6\columnwidth]{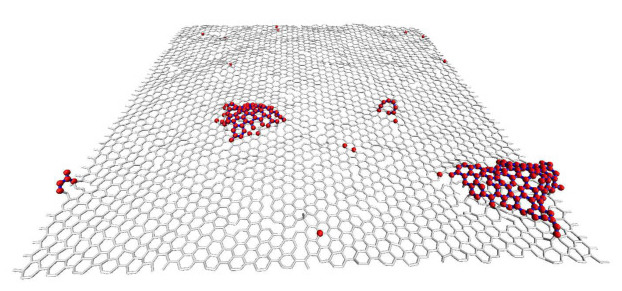}}&
\subfloat[2000K, gas]{\includegraphics[width=0.3\columnwidth]{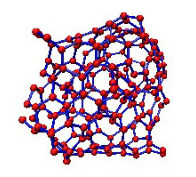}}\\
\subfloat[2500K, surface, 498 ns]{\includegraphics[width=0.6\columnwidth]{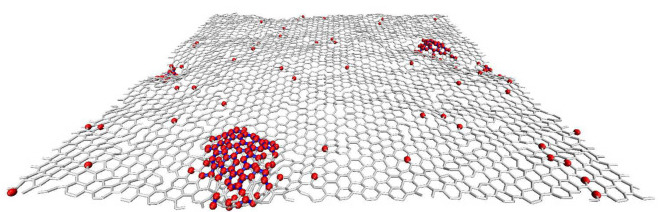}}&
\subfloat[2500K, gas]{\includegraphics[width=0.3\columnwidth]{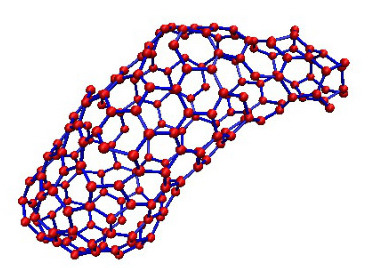}}&
\subfloat[3000K, surface, 539 ns]{\includegraphics[width=0.6\columnwidth]{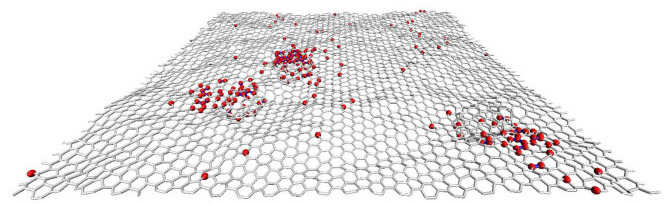}}&
\subfloat[3000K, gas]{\includegraphics[width=0.3\columnwidth]{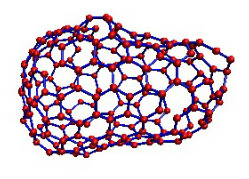}}
\end{tabular}
\caption[Cluster formation on the surface and in the gas phase after long times]{Cluster formation on the surface and in the gas phase after long times (0.5 - 1 $\mu$s). The results are similar to Figure~\ref{fig:snap100} as there are carbon chains at 500K on the surface and in the gas, irregular clusters at 1000K and 1500K, and regular cages at 2000K, 2500K and 3000K.}
\label{fig:snaplong}
\end{figure*}

The next set of results show the cluster formation at longer times, approaching 1 $\mu$s (Figure~\ref{fig:snaplong}). The calculations run faster at lower temperatures so the results for 500K are after 915 ns but at 2500K, the results are shown after 500 ns. They show the continued growth in complexity of molecular structures from 100 ns. 

\begin{figure}
\begin{tabular}{cc}
\subfloat[2500K]{\includegraphics[width=0.45\columnwidth]{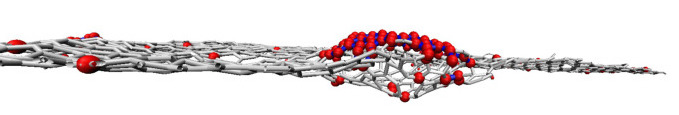}}&
\subfloat[3000K]{\includegraphics[width=0.45\columnwidth]{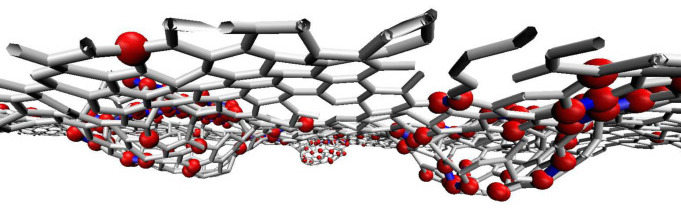}}
\end{tabular}
\caption[VMD snapshots: spherical molecules at 2500K and 3000K after long times]{Cluster formation on the surface at 2500K and 3000K after 498 ns and 539 ns respectively. (a) is similar to Figure~\ref{fig:snapsph100}, showing how after longer times, carbon cages can form on the surface at lower temperatures.}
\label{fig:snapsphlong}
\end{figure}

At 500K there is little change on the surface or in the gas phase as the large chains remain morphologically unaltered. At T $=$ 1000K, the molecules look different on the surface and in the gas: the clusters on the surface are still isolated and irregular but in the gas phase, they are more regular and spherical in shape. In the gas phase at 2000K, 2500K and 3000K, there are regular carbon cages which resemble fullerene-like molecules. At 2000K and 2500K on the surface, the clusters have arranged themselves into regular graphene-like flakes, lying on the surface and starting to deform the surface into spherical shapes, similar to the behaviour which occurred after 100 ns at 3000K. After 539 ns, there are now three cavities in the surface at 3000K as seen in Figure~\ref{fig:snapsphlong}.

We are now seeing fullerene-like molecules forming at later times on surfaces at 2000K and 2500K.  This is at higher temperatures than the observed grain surfaces in Tc-1, which are at 180K - 330K. Our results imply one of two things: either that, given a long enough time, molecules can form curvature and cage-like structures on the cold surfaces; or that cage-like structures must form in the gas phase before depositing onto the surface. The latter is more likely as we have only been able to simulate the formation of flat molecules at low temperatures but cage-like structures can form in the gas at 1000K, similar to the conditions in the gaseous environment of Tc-1.

We are confident that these simulations are representative of typical runs. The results presented are in accordance with the formation pathways seen during preliminary testing of the simulation code and furthermore, we have been able to reproduce the work of \citep{patra2014nucleation} for carbon molecule formation in the gas phase, despite having a difference in density of a few orders of magnitude. We find long carbon chains at low temperatures, irregularly shaped molecules at intermediate temperatures and carbon cages at the highest temperatures.

\subsection{Nucleation time}
\label{sec:nuctime}

The next aim of our work is to determine the nucleation time of carbon-rich clusters on a graphitic surface and compare this to the behaviour in the gas phase. To do this, it is necessary to look at significantly shorter timescales than in Section~\ref{sec:morph} (typically $<$ 1 ns). On longer timescales ($>$ 5 ns), the large clusters have already formed and do not change much over time except for local, small scale evolution. We must hence look at the earliest times to see how the small clusters form. Again, we use 200 carbon atoms in the simulations and look at results over the first nanosecond. In real astronomical environments, where the density is lower, the timescales would of course be much longer because the mean free path between atoms is large.

\begin{figure*}
\includegraphics[width=1.4\columnwidth]{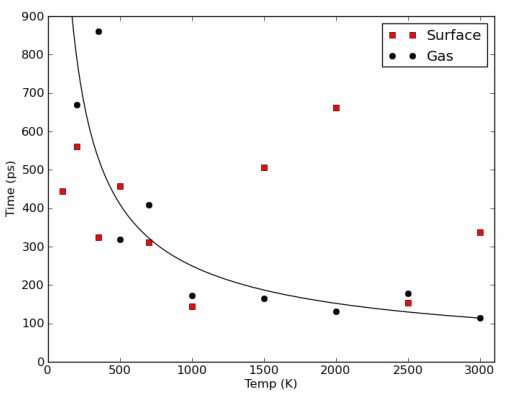}
\caption[Nucleation time for a thirty atom cluster in the gas phase and on the surface as a function of temperature]{Nucleation time for a thirty atom cluster in the gas phase (black dots) and on the surface (red squares) as a function of temperature. For the formation in the gas, the relationship can be modelled with a $t(ns) = 34.942~T^{-0.715} (K)$ power law (solid line). This is similar in behaviour to the results of \citep{patra2014nucleation} with a $t(ns) = 23,153~T^{-1.1} (K)$ power law. The sample size is smaller than used in \citep{patra2014nucleation} , as there are a large number of surface C atoms; therefore the shorter time scale. At low temperatures, the formation is generally quicker on the surface than in the gas but above 1000K this breaks down as atoms have enough energy to interact with the surface.}
\label{fig:nuctime}
\end{figure*}

Figure~\ref{fig:nuctime} shows the nucleation time of a 30-atom cluster in the gas phase and on graphene, with temperature. Each point has an error of $\pm$1 ps due to the fact that particle trajectories were visualised in VMD after each picosecond. In the gas phase, the relationship follows a power law with $t(ns) = 34.942~T^{-0.715} (K)$, found using a least squares method \citep{patra2014nucleation}. On the graphene surface, the formation of carbonaceous clusters generally occurs at earlier times than in the gas phase at low temperatures ($\leq$ 1000K). However at higher temperatures, there is no discernible pattern and in most cases, formation is quicker in the gas than on the surface. 

\begin{figure}
\begin{tabular}{c}
\subfloat[200K, 200 ps]{\includegraphics[width=\columnwidth]{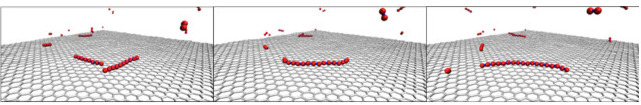}}\\
\subfloat[200K, 500 ps]{\includegraphics[width=\columnwidth]{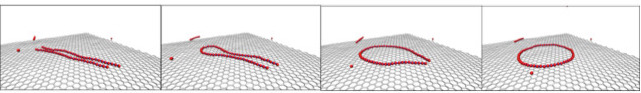}}
\end{tabular}
\caption[VMD snapshots: chains coagulating at 200K]{VMD snapshots at 200K of (a) two short carbon chains joining together after t$\sim$200 ps; and (b) two long chains joining together to form a large carbon ring after t$\sim$500 ps. These illustrate how mobility about the surface at low temperatures catalyses the growth of short chains into larger molecules.}
\label{fig:snapchainscom}
\end{figure}

To explain this behaviour, we must look at the growth of carbonaceous molecules in greater detail. At the lowest temperatures (100-350K), atoms which accrete onto the graphene are suspended about 3.5 \AA ~above the surface as discussed in Section~\ref{sec:morph}. They are free to move around the surface, coalescing to form molecules, chains and larger structures, as depicted in Figure~\ref{fig:snapchainscom}. The surface acts as a catalyst: the molecules are constrained to move only in the x and y directions but their mobility in these directions helps with the growth of large clusters.

\begin{figure}
\begin{center}
\includegraphics[width=\columnwidth]{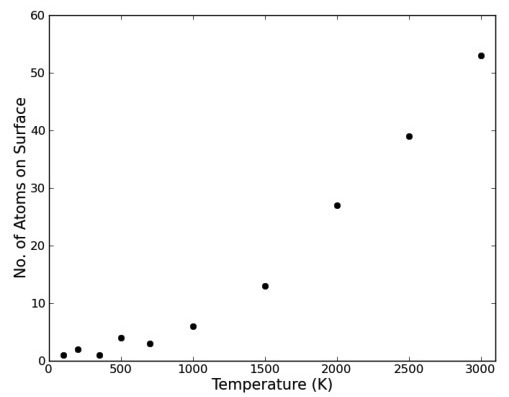}
\caption[The number of atoms bound to the graphene surface after 100 ps over a range of temperatures]{The number of atoms bound to the graphene surface after 100 ps over a range of temperatures. There are few atoms at low temperatures when the atoms are free to move around the surface. The number increases though, and this inhibits cluster growth especially at T $\geq$ 1000K.}
\label{fig:surfatom}
\end{center}
\end{figure}

\begin{figure}
\begin{center}
\includegraphics[width=\columnwidth]{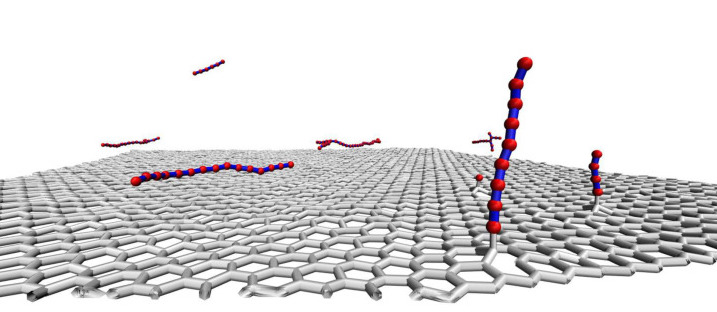}
\caption[VMD snapshots: carbon chains at 500K after 400 ps]{VMD snapshot of carbon chains at 500K after 400 ps: those on the left (mid-ground) are not bound and are hence free to move around the surface; those on the right (foreground) are anchored to the surface.}
\label{fig:snapinter}
\end{center}
\end{figure}

At intermediate temperatures (500-700K), the covalent bonding of gas phase carbon atoms to the surface atoms becomes a determining factor in growth (Figure~\ref{fig:surfatom}). Below 500K, only one or two atoms are bound directly onto the surface and the majority are kept at a distance of 3.5 \AA. Between 500 and 700K, five or six atoms now bind to the surface and we see two classes of carbon chain on the surface: those which are unbound and free to move around; and others which are bound directly to the graphene, as seen in Figure~\ref{fig:snapinter}.

\begin{figure}
\begin{tabular}{c}
\subfloat[2000K, 140 ps]{\includegraphics[width=\columnwidth]{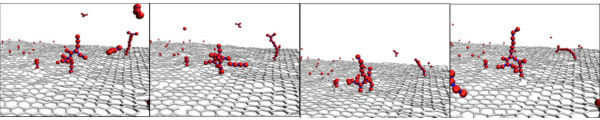}}\\
\subfloat[2000K, 930 ps]{\includegraphics[width=\columnwidth]{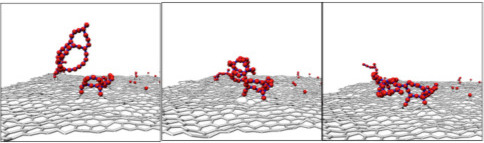}}
\end{tabular}
\caption[VMD snapshots: clusters depositing onto the surface at 2000K]{VMD snapshots at 2000K of (a) a four atom carbon chain in the gas phase bonding to a surface cluster after 140 ps; and (b) a large carbon cluster in the gas phase anchoring to a surface cluster after 930 ps. These show the deposition of molecules in the gas phase onto clusters anchored to the surface.}
\label{fig:snapchainscom2}
\end{figure}

Beyond this, the number of atoms which bind to the surface increases almost linearly with temperature, see Figure~\ref{fig:surfatom}. This has the effect of inhibiting the growth of carbon clusters at higher temperatures ($\geq$1000K) as atoms cannot move around the surface: they are anchored to one spot. For a cluster to grow on the surface, it must first form in the gas phase before binding onto the graphene (Figure~\ref{fig:snapchainscom2}). The dominance of surface binding at high temperatures explains the fluctuations in the nucleation time as molecular growth on the surface now depends on the behaviour in the gas phase. We therefore conclude that mobility on the surface at low temperatures enhances the rate of molecular complexity and growth. Conversely, sticking on the surface at high temperatures hinders the evolution of clusters.

\subsection{Size distributions}
\label{sec:popdist}

We now investigate the size distributions of carbon clusters which grow on the surface, by binning into groups according to the number of atoms in the clusters. To this end, we will again look at short timescales ($<$ 1 ns) as after long times there is little change in the cluster size and the most interesting growth occurs at the beginning. We will use 512 atoms rather than 200 atoms to give better statistics.

\begin{figure*}
\caption[Population distribution for clusters forming in the gas phase from 100K to 3000K]{Size distribution histograms for clusters forming in the gas phase (a) and on the graphene surface (b), from 100K to 3000K. There are 512 C-atoms in the gas and 4128 C-atoms forming the graphene surface. The blue distributions are all found after 0.25 ns. The red distributions are from earlier times and show that at all temperatures, a gamma distribution can be found.}
\begin{tabular}{c}

\includegraphics[width=1.6\columnwidth]{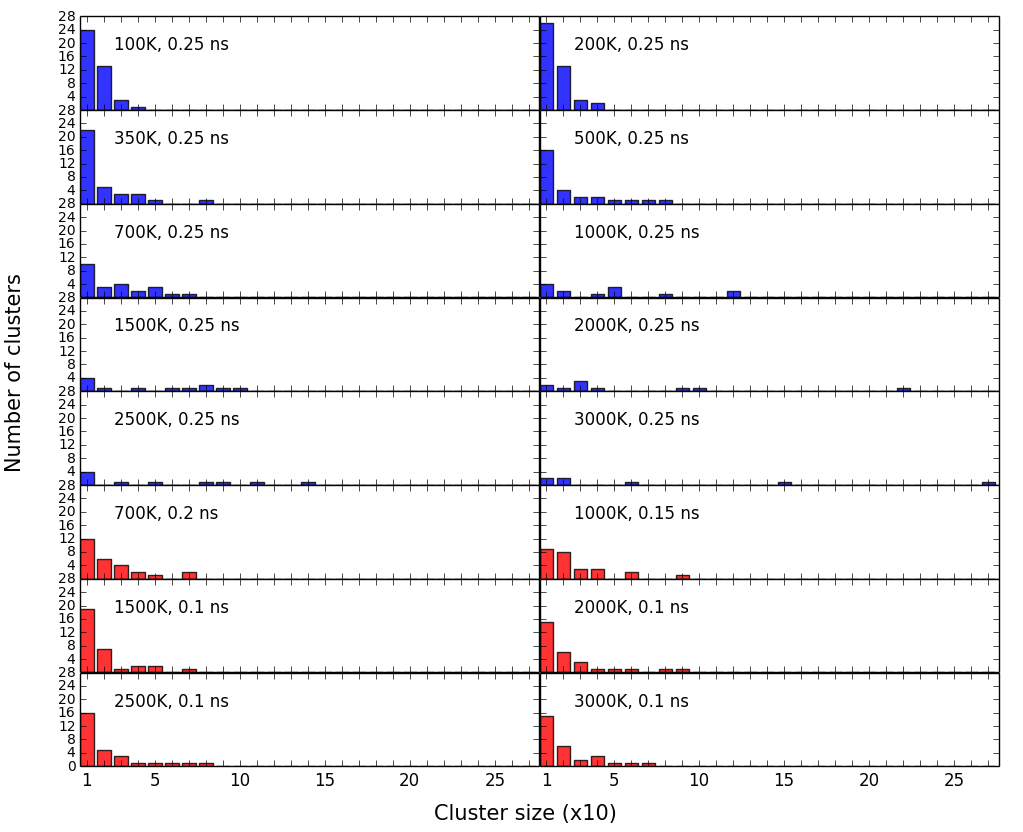}(a)\\
\includegraphics[width=1.6\columnwidth]{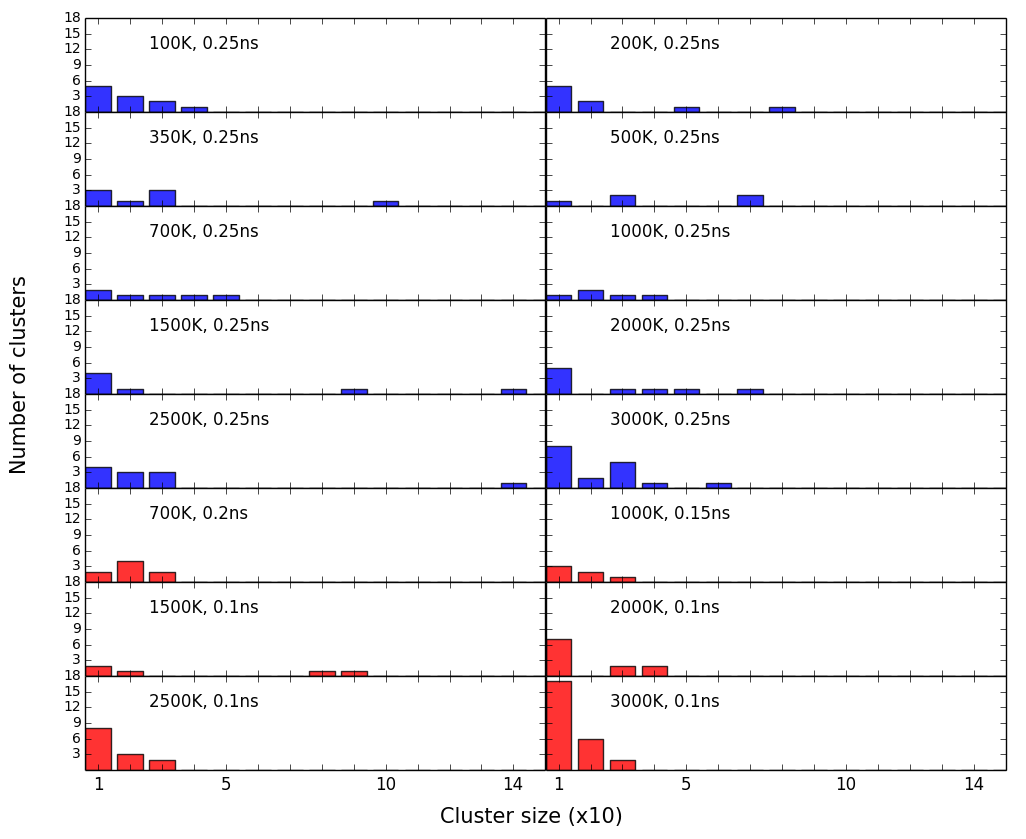}(b)

\label{fig:popdistgas}
\end{tabular}

\end{figure*}

The size distribution histograms at temperatures from 100K to 3000K at times between 0.1 ns and 0.25 ns are shown for the gas phase and on the surface in Figure~\ref{fig:popdistgas}(a) and \ref{fig:popdistgas}(b), respectively.

In the gas phase, we are able to find the same regular distribution shape across all temperatures. There are a few minor differences between them but they can all be said to exhibit the hierarchical growth of a large population of small chains into a few longer chains. They take a form which is similar to a Gamma probability density distribution, as used by \citep{patra2014nucleation}. This is a two parameter function which exhibits an exponential or a chi-squared distribution. The exponential form (with k = 1, $\theta$ = 2; or k = 0.5, $\theta$ = 1) is relevant here for the formation of clusters in the gas phase. The probability distribution function is given by:

\begin{equation}
P(x) =\Bigg( \frac{1}{\Gamma (k) \theta^{k}}\Bigg) x^{k-1}e^{-\frac{x}{\theta}}
\label{eq:gamma}
\end{equation}
where k and $\theta$ are scaling parameters and $\Gamma (k)$ is a Gamma function of argument k.  

In Figure~\ref{fig:popdistgas}(a), at low temperatures (100K, 200K, 350K and 500K), the distribution more closely resembles a Gamma distribution, after 0.25 ns, while at this elapsed time at higher temperatures, the distributions are spread out towards the largest bins. The biggest cluster contains 269 atoms at 3000K. By looking at earlier times though (red distributions), we can find similar exponential-like distributions at the highest temperatures and see how they evolve into their blue counterparts at later times: the size distributions start with many small clusters (occupying the first bin) and evolve over time by coagulation to form a few large clusters (occupying bigger bins).

For the size distributions on the surface (Figure~\ref{fig:popdistgas}(b)), we do not find the same Gamma distribution. To an extent, the same distribution can be seen at 100K after 0.25 ns where comparable sized clusters form on the surface and in the gas. At 200K and 350K though, larger clusters form on the surface than in the gas: 79 atoms at 200K and 92 atoms at 350K on the surface; and 32 atoms at 200K and 59 atoms at 350K in the gas phase. The mobility above the surface at low temperatures helps to form large carbon chains in the same way that it increased the formation rate in Section~\ref{sec:nuctime}. Clusters of comparable size form at 500K on the surface and in the gas.

At earlier times- 0.1-0.2 ns- the distributions of molecular clusters, binned in sizes of ten, follow a Gamma distribution, at all temperatures, but at later times in the simulation, at low temperatures where the molecules formed in the gas phase can have mobility on the surface, the size distributions maintain the Gamma distribution, losing the distinct behavior with increasing temperatures, 1000-3000 K. This is due to the fact that increased kinetic energy prevents interaction with the surface, disturbing the Gamma distribution.



These results show that mobility on the surface helps to grow large carbon chains at low temperatures but large clusters can form in the hotter gas and deposit onto the surface.

\subsection{Temperature gradient between surface and gas atoms}
\label{sec:tempgrad}

So far, we have only studied cases where the surface and gas atoms are kept at the same temperature, but this need not be true in real astrophysical conditions. As previously mentioned, the fullerene spectral data collected by \citep{cami2010detection} suggests that in the environment of the planetary nebula Tc-1, the dust grains are at T $\sim300$K whereas the surrounding gas is at T $\sim1000$K. It is also possible for small grains to be heated by thermal radiation to temperatures that are hotter than the local environment \citep{lequeux2005interstellar}. This effect though, only applies to the smallest grains as their size needs to be as the absorbed wavelength for significant heating to occur. The first case is therefore more likely to be seen in astronomical environments but we can nevertheless examine both ideas.

\begin{figure*}
\begin{tabular}{ccc}
\subfloat[Surface at 300K; gas at 300K]{\includegraphics[width=0.6\columnwidth]{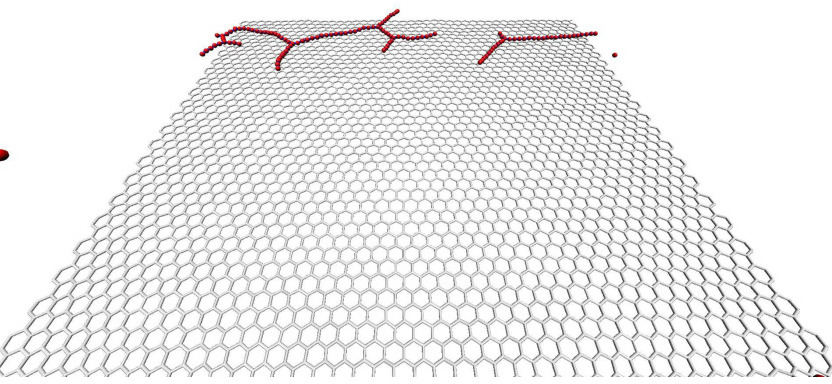}}&
\subfloat[Surface at 300K; gas at 500K]{\includegraphics[width=0.6\columnwidth]{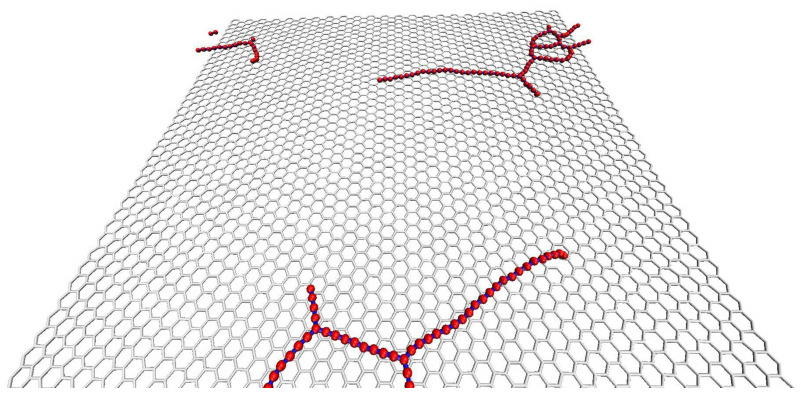}}&
\subfloat[Surface at 300K; gas at 1000K]{\includegraphics[width=0.6\columnwidth]{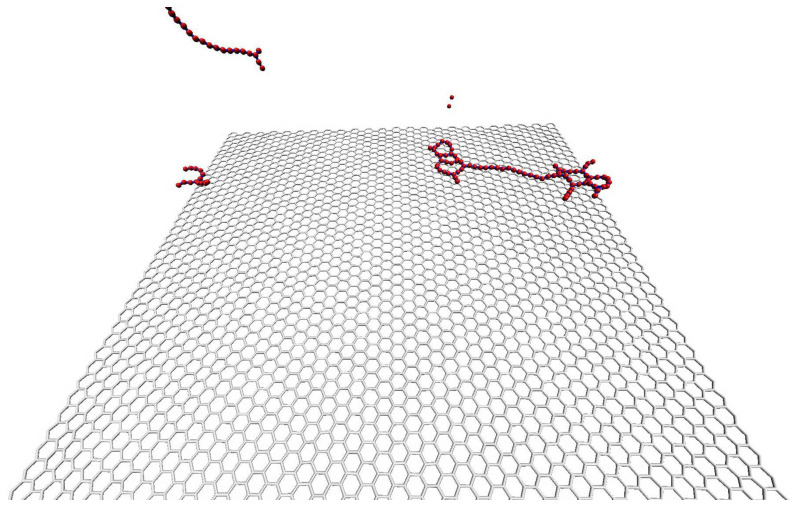}}\\
\subfloat[Surface at 500K; gas at 300K]{\includegraphics[width=0.6\columnwidth]{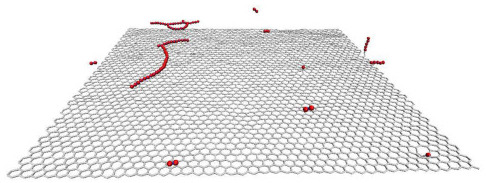}}&
\subfloat[Surface at 1000K; gas at 500K]{\includegraphics[width=0.6\columnwidth]{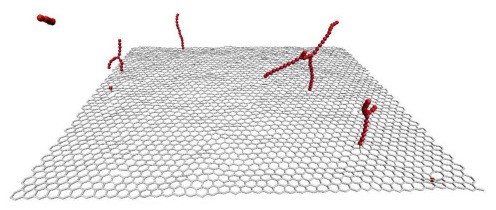}}
\end{tabular}
\caption[VMD snapshots: cluster formation with a temperature gradient]{Snapshots of cluster formation with a temperature gradient between the surface and gas atoms. There are long carbon chains on the surface at 300K for all gas temperatures and this is representative of surfaces at 50K and 100K. When the surface is at 500K and 1000K, we see isolated clusters forming on the surface.}
\label{fig:snapgrad}
\end{figure*}

Our simulations are run with the surface thermostatted at 50, 100, 300, 500 and 1000K, and the gas atoms at 300, 500 and 1000K. The same simulation box is used as before and starts with 200 atoms in the gas. The calculations are performed over 50 ns. At low surface temperatures (50-300K), there is a stark bimodal distribution regardless of the conditions of the gas phase. There is one smaller cluster (3-4 atoms at 50K; 5-39 atoms at 100K; 5-48 atoms at 300K) and one larger cluster (108-130 atoms at 50K; 90-154 atoms at 100K; 97-123 atoms at 300K). These take the form of chains with broadly the same morphology, although there are more carbon loops at higher gas temperatures as shown in Figure~\ref{fig:snapgrad}. At these temperatures, the gas atoms do not interact with the surface at all, so the long carbon chains are made by the coagulation of shorter ones as described earlier.

At higher surface temperatures (500 and 1000K), the largest cluster is made of 65 atoms when the gas is at 1000K. Most clusters fall into the first bin with less than 25 atoms. As the thermal energy of the surface increases, gas atoms can readily bond onto it creating isolated clusters which do not interact with each other.

Notably, in simulations where the surface is hotter than the gas atoms, clusters only form which contain less than 50 atoms but when the surface is cooler, over 100 atoms can join together to form a molecule. The latter condition more resembles astrophysical environments. For all three gas atom temperatures, the size of the cluster that can be made after one ns decreases as the surface temperature increases; 
for T=500K and 1000K, the decrease is present but not as steep.

Broadly speaking, for a surface at a given temperature, the distributions for each different gas temperature are the same. This is certainly true when the surface is at 50K and generally the case for the other temperatures, $\pm$ 1 bin. This suggests that the condition of the surface is the main contributing factor to the morphology of the forming molecules. It also implies that if we want to see a fullerene-like structure on a cold surface, it must form in the gas phase first as the results here suggest that only flat molecules can form on a cold surface.


\subsection{Graphene layers as a graphite analogue} 
\label{sec:lay}

Our simulations are performed with a surface made from one atomic layer of graphite - graphene - which is not realistic. The more analogous example will be graphite which is structurally made up of layers of graphene and has been observed in space \citep{tielens2011chemical}. 

The layers in graphite can be stacked in a few different ways but the most common and lowest energy configuration is ABAB-stacking, where the first and second layers are offset slightly so that the hexagons in the honeycomb lattice do not align \citep{charlier1994graphite}. As a result, some atoms have neighbours above and below them ($\alpha$ atoms) while the others are above and below the empty hexagonal lattice gaps ($\beta$ atoms). AA-stacking describes the arrangement where hexagons in each layer align exactly, but this configuration is less energetically favourable than the ABAB-stacking and has a larger interlayer separation.

While graphite is still a durable material, it does not exhibit the same exceptional properties as graphene. The melting point of graphite has been experimentally measured in the range 4000 - 5000K but free energy calculations put the figure at approximately 4250K \citep{colonna2009properties}. Although the layers in graphite have the same strong covalent sp$^{2}$ bonds as graphene, the interlayer interactions are mediated by far weaker overlapping p$_{z}$ orbitals \citep{schabel1992energetics}. While the bonding in the layers has an energy of 5.9 eV, between the layers its energy is only 50 meV \citep{schabel1992energetics}.

The weak binding of graphite layers is thought to mainly originate from attractive van der Waals interactions ($\frac{1}{r^{6}}$) due to dipole-dipole interaction of p-electrons in the carbon orbitals and a repulsive ($\frac{1}{r^{12}}$) term from the Pauli Exclusion Principle \citep{charlier1994graphite}. This inter-planar distance has an equilibrium value of 3.35\AA~\citep{baskin1955lattice}. There is also a contribution from electronic delocalisation caused by an increase in charge between the layers as electrons are transferred from $\alpha$ to $\beta$ atoms \citep{charlier1994graphite}. This also explains why there is a smaller gap in the ABAB-stacking than for the AA-stacking as all the atoms in the latter case are $\alpha$ atoms so there is no electron transfer. By investigating molecule nucleation with graphene layering, we can glean additional information regarding the effect of extra van der Waals influences from multiple layers on the formation of organic molecules.

In order to create the multilayer graphene, we start by making single sheets in the simulation box. The system is then relaxed for about one ns as the layers come together. This successfully produces multilayers of graphene in an ABAB arrangement (as energetically preferred by graphite) separated by an average distance of 3.38$\pm$0.04 \AA, which is satisfactorily close to the experimental value of 3.35 \AA. This suggests that our method can reproduce the van der Waals interlayer interactions in graphite, through the Leonard-Jones term (Equation~\ref{eq:lj}).

\begin{figure}
\begin{tabular}{cccc}
\subfloat[Monolayer]{\includegraphics[width=0.47\columnwidth]{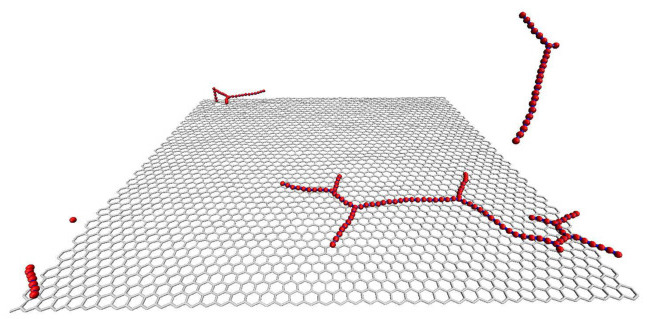}}&
\subfloat[Bilayer]{\includegraphics[width=0.47\columnwidth]{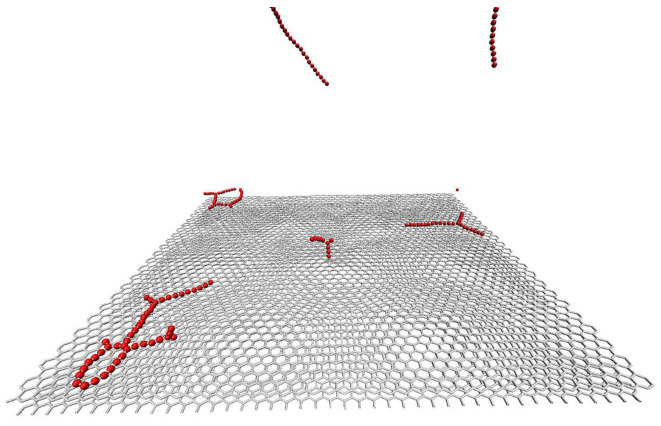}}\\
\subfloat[Trilayer]{\includegraphics[width=0.47\columnwidth]{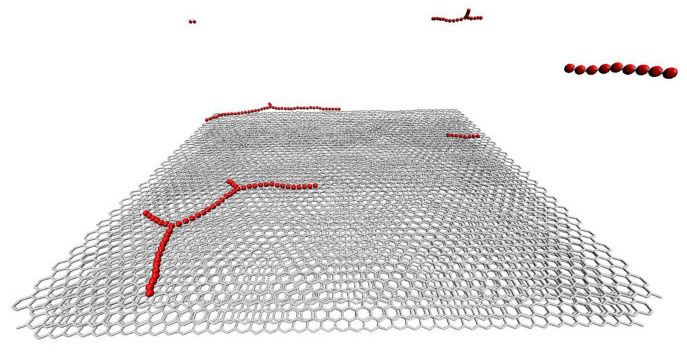}}&
\subfloat[5-layer]{\includegraphics[width=0.47\columnwidth]{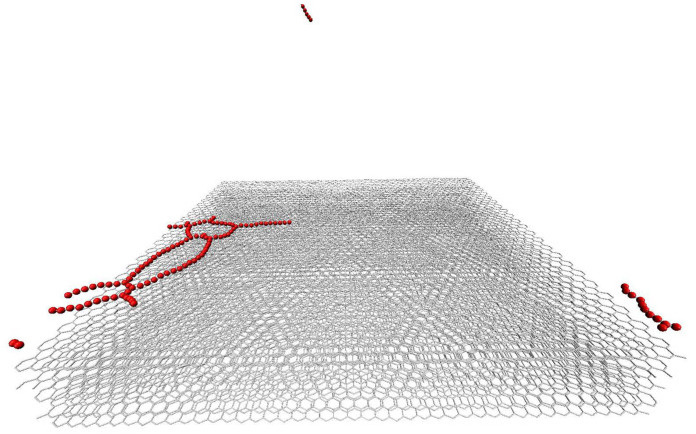}}
\end{tabular}
\caption[VMD snapshots: Cluster formation on graphene with 1, 2, 3 and 5 layers after 1.5ns]{Cluster formation on graphene with 1, 2, 3 and 5 layers after 1.5ns. Branched carbon chains form in all cases.}
\label{fig:snaplay}
\end{figure}

We can then run the simulations over 5 ns at 300K with 200 atoms and utilise a surface with 1 (monolayer), 2 (bilayer), 3 (trilayer) and 5 (5-layer) graphene sheets stacked on top of each other. We then compare the resulting distributions and morphologies to see if there are any significant differences.  Figure~\ref{fig:snaplay} shows snapshots taken after 1.5 ns of the different surfaces.  The cluster evolution for each surface is broadly the same, and can be modelled as a gamma distribution spreading out to larger cluster sizes. This seems to happen at the same rate too, as the cluster size for each case at each time is comparable, although it may be argued that growth is quicker on the monolayer as it plays host to the largest cluster at most intervals.

As the carbon formation appears to be generally similar irrespective of the number of layers in the surface, we can conclude that using a single layer of graphene in our simulations is justified as a means to emulate the interactions during deposition onto carbonaceous astronomical dust grains.

\subsection{Carbon inflow}
\label{sec:inf}

Violent events such as accretion, cloud collisions, supernova explosions and stellar outflows, perturb and enrich the interstellar medium frequently in the form of interstellar shocks \citep{draine1993theory}. This causes an increase in pressure of the local environment, which then expands and develops a shock front at the edge of the expansion. This front can move at supersonic speeds depending on the severity of the initial disturbance. Shock speeds from the most extreme supernova explosions can reach as much as 3000 km/s \citep{nikolic2013integral}. Shocks which occur in dense media like molecular clouds are a couple of order magnitudes slower and models for the production of SiO in molecular outflows have looked at shock velocities of 10 - 50 km/s \citep{gusdorf2008sio, gusdorf2008sio2}. 

\begin{figure*}
\begin{tabular}{ccc}
\subfloat[10 km/s, 14 ps]{\includegraphics[width=0.65\columnwidth]{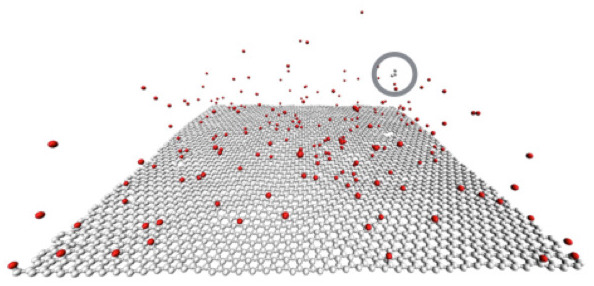}}&
\subfloat[10 km/s, 50 ps]{\includegraphics[width=0.65\columnwidth]{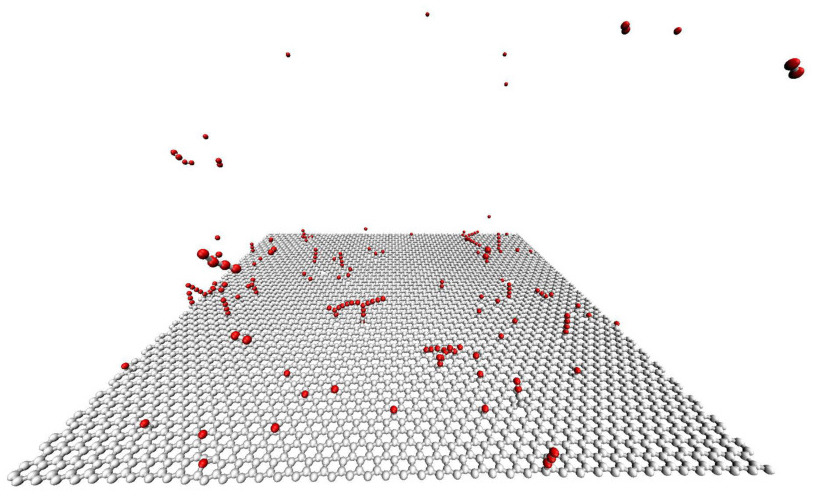}}&
\subfloat[10 km/s, 500 ps]{\includegraphics[width=0.65\columnwidth]{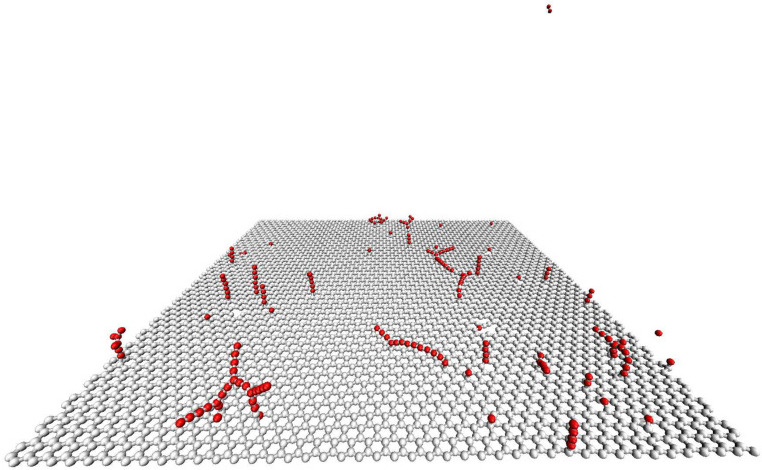}}\\
\subfloat[25 km/s, 0.5 ps]{\includegraphics[width=0.65\columnwidth]{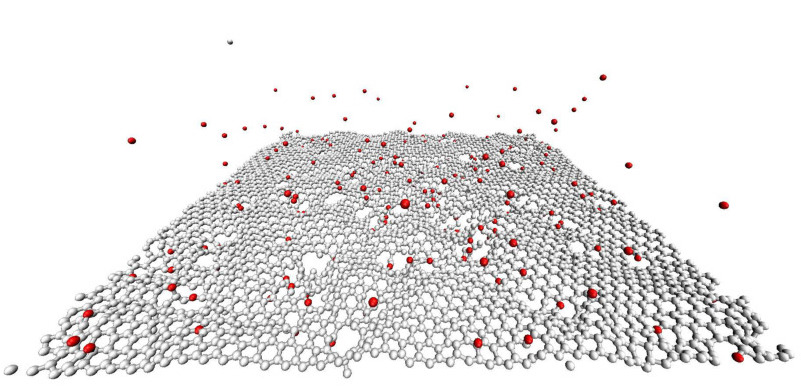}}&
\subfloat[25 km/s, 50 ps]{\includegraphics[width=0.65\columnwidth]{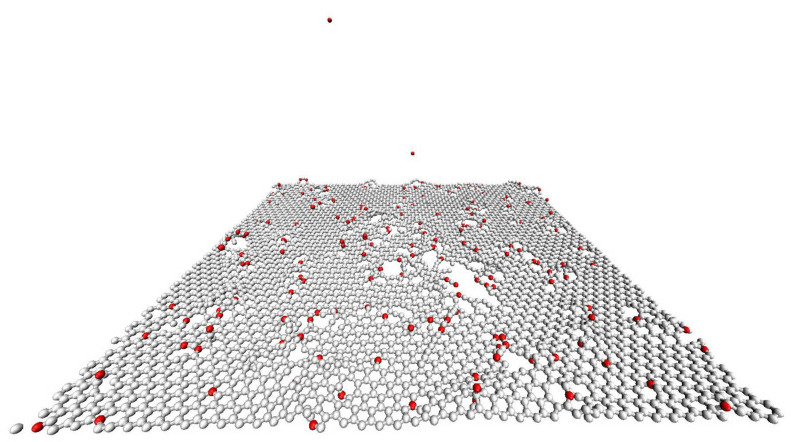}}&
\subfloat[25 km/s, 500 ps]{\includegraphics[width=0.65\columnwidth]{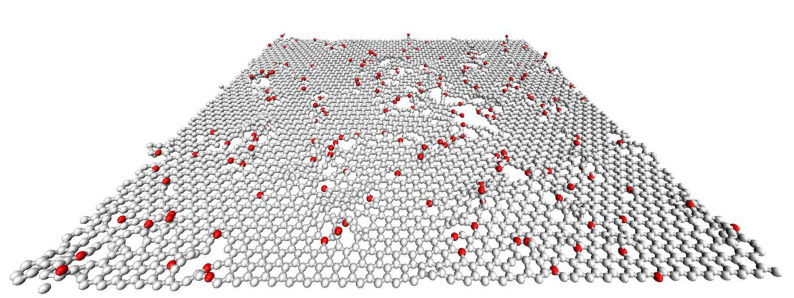}}\\
\subfloat[35 km/s, 0.5 ps]{\includegraphics[width=0.65\columnwidth]{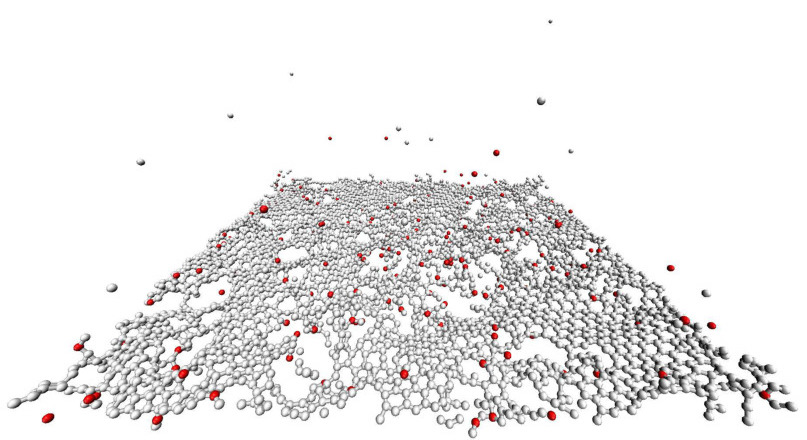}}&
\subfloat[35 km/s, 25 ps]{\includegraphics[width=0.65\columnwidth]{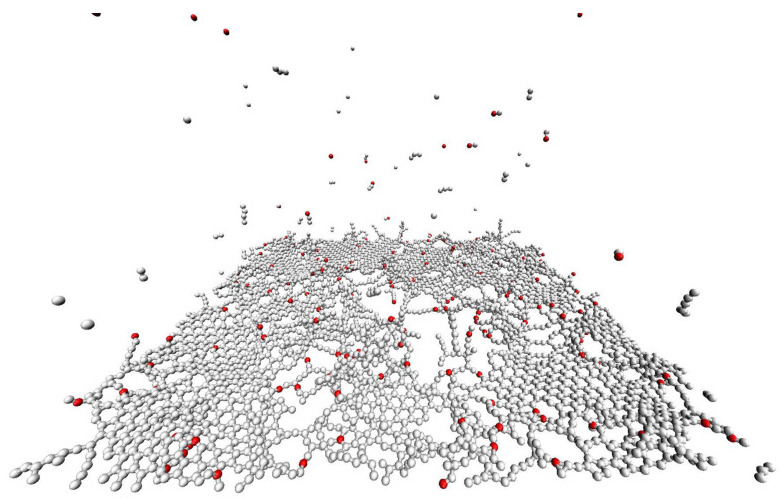}}&
\subfloat[35 km/s, 500 ps]{\includegraphics[width=0.65\columnwidth]{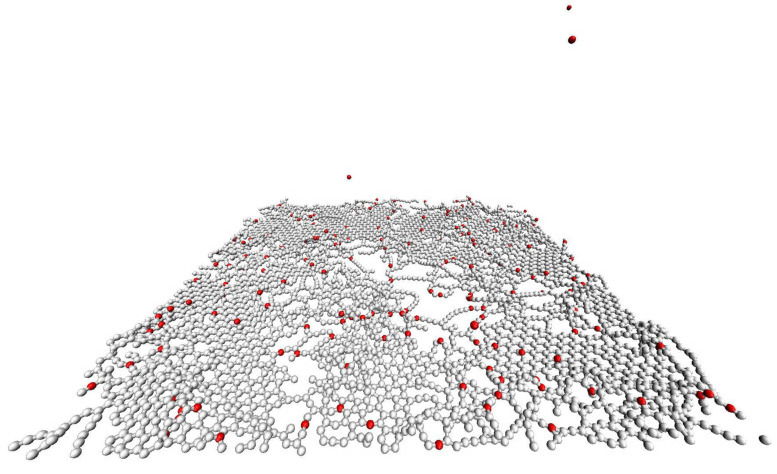}}
\end{tabular}
\caption[VMD snapshots: Effects of an inflow of carbon atoms onto a carbonaceous surface at interstellar shock speeds in the range 10-35 km/s]{Effects of an inflow of carbon atoms onto a carbonaceous surface at interstellar shock speeds in the range 10-35 km/s. At 10 km/s the surface is unaffected by the inflow and small chains can bind to the surface and grow over time. A single triatomic is ejected from the surface (circled). At 25 km/s, the surface is punctured by the incoming carbon atoms and some surface atoms are dislodged. The graphene still has a regular structure though. At 35 km/s, the surface is torn apart and many atoms are ejected. The graphene loses its shape and becomes amorphous.}
\label{fig:snapinf}
\end{figure*}

Previous simulations had looked at a steady state where atoms deposit onto the surface gradually over time (Sections~\ref{sec:morph}-\ref{sec:popdist}). In this section, we will include an inflow to simulate the effects of interstellar shocks. Figure~\ref{fig:snapinf} shows the snapshots of the inflow of carbon atoms onto the surface at shock speeds ranging from 10 - 35 km/s over a short time. When the atoms approach at 10 km/s, the surface is relatively unaffected and the majority of inflowing atoms simply bounce off the surface. They do not have enough kinetic energy to break the strong carbon bonds in graphene. A single triatomic molecule is thrown up from the surface (Figure~\ref{fig:snapinf} (a), circlred) but it later returns to the surface, as does the rest of the material. Over time, the atoms come together to form isolated clusters on the surface in a similar way to results seen in sections~\ref{sec:morph},~\ref{sec:nuctime} and~\ref{sec:popdist}. Our earlier findings are hence reflective of low turbulence conditions.

As the incoming velocity is increased to 25 km/s, the atoms now have more energy and puncture the surface in several places. Where this occurs, the local area loses its shape, creating voids and throwing diatomic and triatomic molecules into the space above, where they can coagulate with other atoms to form short chains. This shows a process whereby molecules could potentially form on a surface and later be ejected into the ISM by collisional shocks. Despite these collisions, the graphene surface retains its global regular hexagonal structure. 

At 35 km/s, the surface is torn apart, losing most of its shape and becoming amorphous. As before, monatomic, diatomic and triatomic molecules are ejected from the surface as a result of the collisions but this time, they are far more numerous in number. Over time, these loose atoms are absorbed back into the amorphous structure-less surface.

We have only simulated the first 0.5 ns, but there is little change in the surface after about 0.2 ns. Together these results suggest a processing method whereby the surface can help large carbonaceous molecules form in low turbulent environments (as in earlier sections at low temperatures) to be later released into the local environment by more tumultuous events.

Strong shocks in warm interstellar medium phases result in grain processing through thermal sputtering, inertial sputtering, grain-grain collisions and vaporisation \citep{tielens2005physics}. These mechanisms release material from grains into the gas phase and column densities for these species can give information on shock speeds and the local environment. Inertial sputtering of gas atoms on grains, as investigated here, disrupts the grain mantle at low velocities and the grain core at high velocities.  Models by \citep{may2000sputtering} show a sharp increase in atomic sputtering fractions between 25 and 45 km/s, and also predict the complete destruction of grains at 50 km/s. Extrapolation of our results support this prediction, as higher incident velocities would further break down the graphene surface.


\section{Conclusion}

We have investigated the formation of carbon-rich molecules and clusters on a graphitic surface as an analogue to an astrophysical surface using reactive molecular dynamic simulations. As discussed above, we find that at low temperatures (100K - 500K) large chains and branched molecules form on the surface over long timescales (100 ns). They do not have enough energy to interact with the surface, and freely move around at a distance of 3.5 \AA. Above 1000K, the carbon atoms can bond to the surface and so form isolated clusters, anchored to one spot. At the highest temperatures, these clusters can evolve into cavities that resemble fullerene-like molecules. 

By looking at times less than one ns, we have been able to determine the nucleation time for carbon molecules and find that the surface helps to catalyse faster growth than in the gas phase at low temperatures, but at higher temperatures this trend breaks down due to surface interactions which prevent short chains from coagulating together, hindering growth. Mobility about the surface at low temperatures also results in larger chains forming than in the gas phase.

We have also shown that the temperature of the surface is a major contributing factor to the molecular structure that forms as similar distributions can be made when the gas particles are at different temperatures for the same surface. As the surface temperature increases, the size of the molecules that can form decreases for all gas temperatures. Furthermore, when the surface is cooler than the surrounding gas, as is typical in astrophysical environments, larger molecules form than when it is hotter.

In addition, the velocity of the incoming gas particles greatly affects the formation of carbon molecules on the surface. At low interstellar shock speeds (10 km/s), isolated clusters can still form on the graphene but at higher speeds, the surface starts to deform and become more amorphous so that any new particles simply integrate themselves into the dendritic structure. These higher velocities do show how molecules can be ejected from the surface through collisions, for reprocessing in the gas above.

In conclusion, this work provides new insights into the formation of large carbon-rich molecules on astrophysical carbonaceous surfaces. While the particle densities considered in the current simulations are not nearly as tenuous as those in the ISM, the results presented in this work are at the threshold of lowest feasible densities and temperatures and are certainly done at the longest time scales. The higher densities make the simulations feasible and can therefore point to processes occurring under realistic astrophysical conditions. This method could be extended to study molecular formation on dust grain surfaces beyond the idealised graphene sheet that was employed here. One future step is to study formation on silicate surfaces. Chemical reactions involving carbon-chain and PAH molecules, at low temperatures, would necessarily have to involve quantum tunnelling. The Laboratory experiments will further reveal the grain-surface chemistry of interstellar ices and bring us closer to understanding the origins of the prebiotic molecules that lead to life on Earth.

\section*{Acknowledgements}

DWM acknowledges the financial support from the University of Southampton Bursary Fund and the NSF through a grant to the Institute for Theoretical Atomic Molecular and Optical Physics. The authors thank the anonymous referee for many valuable suggestions. While this acknowledgement may be unusual, we feel that it's warranted, as the referee should be credited for concise and precise comments on improving the manuscript.

\bibliographystyle{mnras}
\bibliography{arxiv_submission}

\label{lastpage}
\end{document}